
%
%
%
%
\def\vect#1#2{\setbox0=\hbox{$#1$}   \def\cpy{\copy0\kern-\wd0}
              \def\column{\raise.025em\cpy \raise.0125em\cpy \cpy}
              \def\back{\kern-#2em}  \def\forw{\kern+#2em}
        \hbox{\raise.0250em\cpy      \back\back\column
              \forw\column           \forw\forw\column
              \forw\column           \back\back\box0  } }
%
\global\newcount\meqno
\def\eqn#1#2{\xdef #1{(\secsym\the\meqno)}
\global\advance\meqno by 1 $$#2\eqno#1$$}
%
\global\newcount\refno\def\ref#1#2#3{\global\advance\refno by1
\xdef #1{[\the\refno]}\xdef #2{#3}#1}
%

%
\def\uplrarrow#1{\raise1.5ex\hbox{$\leftrightarrow$}\mkern-16.5mu #1}
%
%
%
\def\ignore#1{}
\def\pair{e^+ e^-}
\def\bbar{\bar B}
\def\u{\Upsilon}
\def\two{{B\bbar}}
\def\twos{{B_s\bbar_s}}
\def\bbpi{{BB\pi}}
\def\three{{B B^\pm \pi^\mp}}

\def\mev{\,{\rm MeV}}
\def\gev{\,{\rm GeV}}
\def\ufr{\u(4{\rm S})}
\def\ufv{\u(5{\rm S})}
\def\usx{\u(6{\rm S})}
%

\def\bstr{{B^*}}
\def\bbst{{\bbar^*}}
\def\ub{\underline{b}}

\def\qbar{\overline q}
\def\feyn#1{/\!\!\! #1} 
\def\vs{\feyn{v}}
\def\projpls{{1+\vs \over 2}}
\def\projmns{{1-\vs \over 2}}
\def\e{\varepsilon}
\def\eb{\vect{\e}{0.015}}
\def\es{\feyn{\e}}
\def\sumeps{{\textstyle\sum\limits_\e}}
\def\bB{{\bf B}}
\def\oB{\overline{\bB}}
\def\uB{{\bf B}\hskip-.715em{\underline{\hphantom n}}\hskip.15em}
\def\ouB{\overline{\uB}}
\def\c{{\cal C}}

\def\fpi{f_\pi}
%

\def\qtwo{Q_{BB}}
\def\qthree{Q_{\three}}
\def\l{{\cal L}}
\def\leff{\l_{\rm eff}}
\def\tr{{\rm tr}}
\def\csb{\Lambda_{\chi}}
\def\amu{A^\mu}
\def\dmu{\partial^\mu}


\def\dk{\Delta k}
\def\k{\vect{k}{0.0125}}
\def\q{\vect{q}{0.0125}}
\def\dkb{\Delta\k}

\def\epi{E_\pi}
\def\tpi{T_\pi}
\def\ppi{p_\pi}
\def\mpi{m_\pi}
\def\ppib{\vect{p}{0.0125}_\pi}
\def\dm{\Delta M}
\def\pp{{\rm PP}}
\def\pv{{\rm PV}}
\def\vv{{\rm VV}}
\def\rpp{r_\pp}
\def\rpv{r_\pv}
\def\rvv{r_\vv}
\def\Rpp{R_\pp}
\def\Rpv{R_\pv}
\def\Rvv{R_\vv}
\def\so{\sigma_0}
\def\sinv{{1\over\so}}
\def\crsc#1{\sigma(\pair\to #1)}
\def\rts{\sqrt{s}}
\def\av{\overline{v}}

\def\ak{\overline{k}}
\def\akb{\overline{\k}}
\def\Ra{R_\alpha}
\def\Rai{\Ra^i}

\def\Rmax{R(s;\emax)}
\def\ra{r_\alpha}

\def\eb{\vect{\e}{0.015}}
\def\a{{\cal A}}
\def\eabcd{\epsilon_{\alpha\beta\gamma\delta}}
\def\acp{A^{\rm CP}}
\def\pk{\psi K_{\rm S}}
\def\et{\e_{\rm tag}}
\def\nb{\,{\rm nb}}
\def\emax{\epi^{\rm max}}
\def\lapprox{\mathrel{\hbox{\setbox0=\hbox{$\sim$}
\lower.8ex\copy0\kern-\wd0\raise.3ex\hbox{$<$}}}}
\def\gapprox{\mathrel{\hbox{\setbox0=\hbox{$\sim$}
\lower.8ex\copy0\kern-\wd0\raise.3ex\hbox{$>$}}}}
\magnification=1200

\vsize=8.9 true in\hsize=6.3 true in
%
\hoffset=0.08 true in
%
\tolerance 10000
\baselineskip 12pt plus 1pt minus 1pt
\centerline{{\bf The Rate for $\vect{\pair\to\three}{0.01}$}}
\centerline{{\bf and its Implications for the Study of CP Violation,
$\vect{B_s}{0.01}$ Identification,}}
\centerline{{\bf  and the Study of $\vect{B}{0.01}$ Meson
Chiral Perturbation Theory}
\footnote{$\star$}{This work is supported in part by funds provided
by the U. S. Department of Energy (D.O.E.) under contract
\#DE-AC02-76ER03069
and in part by the Texas National Research Laboratory Commission
under grant \#RGFY92C6. }}
\bigskip
\smallskip
\def\address{Present address: Department of Physics, University of
Southampton, Highfield, Southampton, SO9 5NH, U.K.}
\def\awards{National Science Foundation Young Investigator Award.\hfill\break
Department of Energy Outstanding Junior Investigator Award.\hfill\break
Alfred~P.~Sloan Foundation Research Fellowship.\hfill\break}
\centerline{Laurent Lellouch,\footnote{$^{\dag}$}{\address}
Lisa Randall\footnote{$^{\ddag}$}{\awards} and Eric Sather}
\bigskip
\centerline{\it Center for Theoretical Physics}
\centerline{\it Laboratory for Nuclear Science and Department of
Physics}
\centerline{\it Massachusetts Institute of Technology}
\centerline{\it Cambridge, MA 02139}
\vskip .15in
\centerline{Submitted to: {\it }}
\vskip .15in
\centerline{\bf Abstract}
{\narrower H.~Yamamoto \ref\yam\yamref{H.~Yamamoto, to be
published.} has proposed employing $B$ mesons produced in
conjunction with a single charged pion at an $\u$ resonance for
studies of CP   violation in the neutral $B$ meson system at a
symmetric $e^+$-$e^-$ collider. The sign of the charged pion would
tag the neutral $B$ meson. We estimate this branching ratio,
employing   the  heavy meson  chiral effective field
theory. We find a negligible branching ratio to $B B^{\pm}
\pi^{\mp}$ at the $\ufv$ and a branching ratio of only a few percent
at the $\usx$. However, if nonresonant studies of neutral $B$ mesons
should prove feasible, Yamamoto's proposal could be a good method
for tagging neutral $B$'s for the study of CP violation at a
symmetric collider.

We also explore the possibility of studying $B_s$ at the $\ufv$.
The rate is low but depends sensitively on the precise value of the
mass of the $B_s$. The  background we compute is comparable to the
rate at the largest allowed value of the $B_s$ mass.

Finally, we discuss the extraction of the axial pion coupling to $B$
mesons from measurement of the $B\bbar\pi$  branching fraction in a
restricted region of phase space, where chiral perturbation theory
should work well.
\smallskip}
\vfill
\vskip -12pt
\noindent CTP\#2155 \hfill {December\ 1992}
\footline={\hfil}
\eject
\pageno=1
\footline={\hss\tenrm\folio\hss}
\baselineskip 24pt plus 2pt minus 2pt
\noindent{\bf 1.\quad Introduction}
\medskip
It is well known that CP violation studies are difficult at a
symmetric collider. In $\pair\to B^0\bbar^0$, the heavy mesons are
produced in a $C$-odd state, so that the time integrated asymmetry
vanishes unless the time ordering of the signal and tag can be
measured. An alternative tagging method for which the time
integrated asymmetry does not vanish has been proposed by
H.~Yamamoto  \yam.  His suggestion is to study  $B$ mesons produced
in conjunction with a single charged pion at an $\u$ resonance, so
that the  sign of the charged pion tags the single neutral $B$ meson
as $B^0$ or $\bbar^0$.  Neutral $B^*$ mesons can also be used, since
they decay immediately via photon emission into pseudoscalars,
before weak mixing or decays have occurred. This method should
provide a simple and efficient tag of the neutral $B$ meson: In
addition to the decay products of the neutral $B$, one only needs to
detect the additional soft pion. The charged $B$ is then tagged by
the invariant mass of the missing four momentum, so it need not be
reconstructed.  Unlike many conventional proposals for the study  of
CP violation in the neutral $B$ system, essentially all the events
are tagged.

The utility of this method depends on the event rate. We calculate
this rate, employing the heavy meson chiral effective theory, in
order to evaluate the potential of Yamamoto's proposal. Although it
is difficult to do a reliable calculation for all relevant kinematic
regions,  we can nevertheless do a calculation which incorporates
propagator enhancements, phase space, and derivative couplings.   We
find that the  branching ratio should be negligible  at the $\ufv$
and  only a few percent at the $\usx$. However, Yamamoto's proposal
could prove  a competitive method for tagging neutral $B$'s for the
study of CP violation at a symmetric collider at slightly higher
center of mass energy, about $12\gev$.

Our calculation is also useful because the $\bbpi$ mode is a
potential background to $B_s$ identification (as discussed in
section~8). We find  the $\bbpi$ background could be comparable to
the rate for $B_s$ production at the largest experimentally allowed
value of the $B_s$ mass, but is most likely small compared to the
rate if the  $B_s$ mass is near the central value of the
experimentally allowed range.

Finally, we discuss the extraction of the axial coupling constant of
the $B$ mesons from a measurement of the $\bbpi$  branching fraction
in a restricted region of phase space, where the heavy meson chiral
lagrangian should apply.

We begin in section~2 by describing the heavy-meson effective
theory, in order to review the assumptions and establish notation.
In section~3, we describe how heavy quark ideas apply to the process
$\pair\to\three$, and in section~4 we construct the effective
lagrangian.  We then calculate the cross section for
$\pair\to\three$ in section~5.  We discuss the implications of our
calculation for CP violation studies  and $B_s$ identification in
sections~6 and 7. In section~8, we discuss the regime of validity of
the results, and the extraction of $g$. Conclusions follow in the
final section.
\bigbreak
\goodbreak
\xdef\secsym{2.}\global\meqno = 1
\noindent{\bf 2. \quad Review of the Heavy Meson Theory}
\noindent
\medskip
In this section, we review the treatment of heavy meson fields in
the heavy meson chiral effective theory. It is convenient to
describe the $\bbar$ and $\bbst$ mesons which contain a bottom quark
of velocity $v$ and a light antiquark, $\qbar$ by a Dirac tensor
field of the form \ref\geo\georef{H.~Georgi, Lectures presented at
the 1991 Theoretical Advanced Study Institute, Boulder (World
Scientific), to be published.}:
\eqn\tens{\bB(v)=\projpls\bigl(-b\gamma_5+\sumeps b_\e \es\bigr).}
Here $b$ and $b_\e$ are the destruction operators for $\bbar$ and
$\bbst$ mesons. The field of the $B$ mesons, which we denote as
$\uB(v)$, can be obtained from $\bB(v)$ by using the
charge-conjugation properties of the $B$ mesons. One obtains
\ref\gri\griref{B.~Grinstein et.\ al.,  Nucl. Phys. B380 (1992) 369.}
\eqn\tensconj{\eqalign{
\uB(v)=&C(\c\bB(v)\c^{-1})^T C^T\cr
=&\bigl(-\ub\gamma_5+\sumeps\ub_\e\es\bigr)\projmns,}}
where $\c$ is the charge conjugation operator and
$C=i\gamma^2\gamma^0$.

We also require fields which create $B$ mesons, obtained from the
destruction fields by Dirac conjugation:
\eqn\Diracconj{\eqalign{\oB(v)&=\gamma^0\bB(v)^{\dagger}\gamma^0\cr
                       \ouB(v)&=\gamma^0\uB(v)^{\dagger}\gamma^0.}}

Under a heavy-quark symmetry transformation the mesons transform as
follows:
\eqn\hqtran{\eqalign{\bB(v)&\to S_v\bB(v),\cr
                   \uB(v)&\to \uB(v)S_v^{\dagger},}}
where $S_v$ is an element of the spin-1/2 representation of the little group
of a particle of velocity  $v$.

We implement the chiral symmetry in the usual way
\ref\cwz\cwzref{S.~Coleman, J.~Wess, and B.~Zumino, Phys.\ Rev.\ 177
(1969) 2239; C.~Callan, S.~Coleman, J.~Wess, and B.~Zumino, 177
(1969) 2247.} by introducing the non-linear field $\xi=e^{i\pi^a
T^a/\fpi}$. Under an ${\rm SU(3)_L\times SU(3)_R}$ chiral
transformation,  $\xi\to L\xi U^{\dagger}=U\xi R^{\dagger}$. This
defines the SU(3) matrix $U$ as a nonlinear function of L, R and
$\pi(x)$.  From $\xi$ one can construct an axial vector field,
\eqn\axial{\amu={i \over 2}
(\xi^{\dagger}\dmu\xi-\xi\dmu\xi^{\dagger})=-\dmu\pi/\fpi+\dots,}
which transforms under the chiral symmetry as $\amu\to   U\amu
U^{\dagger}$. One can also construct a vector field from $\xi$,
\eqn\vector{V^\mu={1\over
2}(\xi^{\dagger}\dmu\xi+\xi\dmu\xi^{\dagger})
=\pi\uplrarrow{\dmu}\pi/2\fpi^2+\dots,}
which functions as a connection term in a covariant derivative:
under a chiral transformation, $\dmu+V^\mu\equiv D^\mu\to U D^\mu
U^{\dagger}.$

According to the usual convention  \ref\wse\wseref{M.~B.~Wise,
Phys.\ Rev.\ D 45 (1992) 2188.}, the field $\bB_a$, which destroys a
$\bbar$ meson containing a light antiquark of flavor $a$, transforms
under the chiral symmetry as
\eqn\chitran{\bB_a(v)\to\bB_a(v)U^{\dagger}_{ab}.}
Similarly, the field $\uB_a$ which destroys a $B$ meson containing
an light quark of flavor~$a$ transforms as
\eqn\conjchitran{\uB_a(v)\to U_{ab}\uB_b(v).}
\bigbreak
\goodbreak
\xdef\secsym{3.}\global\meqno = 1
\noindent{\bf 3.\quad Applicability of the Chiral Heavy Meson Effective Theory
to $\vect{\pair\to\three}{0.01}$}
\medskip
In this section we discuss the energy regimes in which $B$ meson
production can be reliably described using the chiral heavy meson
effective theory. We argue that although such a calculation is not
reliable to better than an order of magnitude within the resonance
region, heavy quark methods should apply at higher center of mass
energy. Because the calculation also requires that the pion energy
be sufficiently  low for chiral perturbation theory to work well,
the calculation with the heavy meson chiral lagrangian is
trustworthy only over a restricted region  of phase space.

In order to become familiar with the energy scales involved in $BB$
and $\bbpi$ production, we begin by listing the masses of the $B$
mesons and the $\u$ resonances and also the total final-state
kinetic energy, or $Q$ value, for $\u$ decay into $B\bbar$ and
$\three$:   The charged pseudoscalar $B$ mesons have mass
$M_B=5278.6\pm2.0\mev$ (all values for particle masses in this
section are taken from ref.\ \ref\pdg\pdgref{Particle Data Group,
K.~Hikasa et. al., Phys.\ Rev.\ D45 (1992).}); the neutral
pseudoscalar mass is essentially the same. The vector $\bstr$ meson
mass is $M_{\bstr}=5324.6\pm2.1\mev$.  In  Table~1 we list the
central values for the masses of the $\u$ resonances and the $Q$
values for decay into two heavy pseudoscalars with and without a
charged pion:
\eqn\qvalue{\eqalign{\qtwo&\equiv M_{\u}-2M_B,\cr
\qthree&\equiv M_{\u}-2M_B-m_{\pi^\pm}.\cr}}
For final states including heavy vector mesons, the $Q$ values are
reduced by one or two times the heavy meson hyperfine splitting,
$M_{\bstr}-M_B=46.0\pm0.6\mev$.  We also list the $Q$ values for a
center of mass energy of $12\gev$, which is above the resonance
regime.

We see that pions produced at a resonance have small momentum  and
energy relative to the chiral scale.  At $12\gev$, the most
energetic pions which are produced can probably not be treated as
soft in a chiral expansion; however, we will see in section~8 that
even at this energy the pion can be treated as soft over a
substantial portion of the phase space.  A naive estimate of the
relative rate of $\three$ to $B\bbar$ production would say that the
first is suppressed by a factor of $(p_\pi/4\pi f_\pi)^2$ relative
to the latter, so that it will only be produced significantly at
center of mass energy approximately $1\gev$ above threshold, which
is why we investigate not only the resonance regime, but higher
center of mass energy as well.
%
\def\row#1#2#3#4
      {&$#1$&&$#2$&&$#3$&&$#4$&\cr}
\def\rowomit{height2pt
      &\omit&&\omit&&\omit&&\omit&\cr}
\def\horizrule{\rowomit\noalign{\hrule}\rowomit}
\topinsert
\centerline{Table 1. Masses and $Q$ Values}
$$\vbox{\offinterlineskip
\hrule\halign{
\vrule#&\quad\hfil#\hfil\quad&\vrule#&\quad\hfil#\hfil\quad&
\vrule#&\quad\hfil#\quad&
\vrule#&\quad\strut\hfil#\quad&\vrule#\cr
\rowomit
\row{\rm Resonance}{\rts\,({\rm MeV})}{\qtwo\,({\rm MeV})}{\qthree\,({\rm
MeV})}
\horizrule
 \row{\u(4{\rm S})}     { 10580 }{  23\qquad }{$---$\qquad }
 \row{\u(5{\rm S})}     { 10865 }{ 308\qquad }{ 168\qquad }
 \row{\u(6{\rm S})}     { 11020 }{ 463\qquad }{ 323\qquad }
\row{\rm Off\ Resonance}{ 12000 }{1443\qquad }{1303\qquad }
\rowomit
\noalign{\hrule}
}}$$
\endinsert

In our calculation, we need both the chiral and heavy quark
approximations to be valid. Some care must be taken in the
application of the heavy quark effective theory to a process with a
$B$ $and$ a $\bbar$ meson in  the final state.  The problem is that
the   Isgur--Wise function (or its analytic continuation)  is not
useful in the resonance regime, where the matrix element $\langle
B(p)\bbar(p')|\overline b \gamma^\mu b|0\rangle$ varies rapidly as a
function of $p \cdot p'$. Because of this rapid variation, the form
factor is not well described in the resonance region as a function
of $v \cdot v'$, where $v$ and $v'$ are the heavy quark velocities.
Furthermore, it is not normalized, and drops rapidly to zero beyond
the resonance region. This behavior was studied by Jaffe in ref.\
\ref\jaf\jafref{R.~L.~Jaffe, Phys.\ Lett.\ B245 (1990) 221.}, and is
expected on the basis of general QCD considerations, since  the
states which can decay into $B$ mesons are not well described as
simple Coulomb bound states.

This has several important consequences. First, in the resonance
region the calculation must be considered as, at best, an order of
magnitude estimate.    In the language of the heavy meson effective
theory, this is because there is no  well-defined derivative
expansion; the rapid variation of the cross section with momentum in
the resonance region is reflected in the heavy meson lagrangian by
the presence of higher derivative operators ``suppressed" only by
the QCD scale.  This is generally not the case in heavy  quark
calculations because of reparameterization
invariance\ref\dug\dugref{M.~J.~Dugan, M.~Golden and B.~Grinstein,
Phys.\ Lett.\ B282 (1992) 142.}\ref\rep\repref{M.~Luke and
A.~V.~Manohar, Phys.\ Lett.\ B286 (1992) 348.}.  However, here,
although the total cross section for $\pair\to\bbpi$ respects a
reparameterization invariance since the cross section depends only
on the $B$ meson momenta ($p$ and $p'$)  and not separately on the
heavy quark velocities ($v$ and  $v'$) and the residual momenta ($k$
and $k'$), the rate for ``decay" of the source is not
reparameterization   invariant. This is because in the effective
lagrangian, we work with heavy quark fields at fixed values of $v$
and $v'$, so that we necessarily assume a resonance produced at
fixed $v\cdot v'$  (not $p \cdot p'$). Therefore, the effective
lagrangian must contain   nonreparameterization invariant derivative
terms to reproduce the full cross section, which does respect
reparameterization invariance.  Since the momentum of a $B$ meson is
of order $\sqrt{T_B M_B}$, where $T_B$ is the kinetic energy of the
$B$ meson,  the higher derivative terms are always large, even in
the heavy quark limit so that  the derivative  expansion is not
reliable.

It is clear that the relative rate predictions for decay to two
pseudoscalars, vector and pseudoscalar, and two vectors, considered
in refs.\ \ref\man\manref{T.~Mannel and Z.~Ryzak, Phys.\ Lett.\ B247
(1990) 412.}, \ref\flk\flkref{A.~F.~Falk and B.~Grinstein, Phys.\
Lett.\ B249 (1990) 314.} and \ref\drj\drjref{A.~De~R\'ujula,
H.~Georgi and S.~Glashow, Phys.~Rev.~Lett.~37 (1976) 398.} are only
appropriate beyond the resonance regime, as has been emphasized by
these authors. The measured \ref\dex\dexref{G.~Goldhaber et.\ al.
Phys.\ Lett.\ B69 (1977) 503.}\ branching ratios for the decay of
the $\psi$(3S) into $D$ mesons of different spins strongly disagree
with those  predicted by a naive application of the results of the
heavy quark theory to the resonance regime
\ref\dgg\dggref{A.~De~R\'ujula, H.~Georgi and S.~Glashow,
Phys.~Rev.~Lett.~38 (1977) 317.}.   The explanation of ref.\
\ref\ley\leyref{A.~Le~Yaouand et.~al., Phys.\ Lett.\ B71 (1977)
397.} is that the nodes in the $\psi$(3S)  momentum-space
wavefunction result in almost no overlap with the final state $D$
mesons except when both are vector particles. In our approach, we
attribute the large discrepancy between the prediction and the
experimental results to  the presence of higher dimension operators,
suppressed only by the QCD scale, which  give different
contributions for decays to heavy mesons of different spins  because
of the significant mass splitting.

We conclude that the only regime where one would trust the
calculation of $B$ meson production  to better than an order of
magnitude is beyond the resonance regime. Fortunately, this is the
region where the result is most interesting since the rate in the
resonance regime is too small. When we calculate in the resonance
region where heavy quark relations are untrustworthy, we view the
calculation as a model which should reproduce important qualitative
features of the true rate. These include the relevant kinematic
features, phase space, propagator enhancements, and derivative
couplings.
\bigbreak
\goodbreak
\xdef\secsym{4.}\global\meqno = 1
\noindent{\bf 4.\quad Effective Theory for $\vect{\pair\to\three}{0.01}$}
\medskip
We proceed to the construction of the effective lagrangian for $BB$
and $\bbpi$ production, keeping in mind  the above caveats regarding
the application of the heavy meson and chiral effective theory to
these processes.

It is useful to divide the range of possible center of mass energies
into three regions: the resonance regime, above the resonance regime
but with the $B$ mesons nonrelativistic, and high energy, with  the
$B$ meson fully relativistic. The heavy quark theory is
straightforward to construct in the third regime, and has been
treated in previous work \man\flk. One couples the current to a
heavy quark and antiquark of velocities $v$ and $v'$. The heavy
quark operator is then evaluated between heavy meson states and the
appropriate spin symmetry relations between amplitudes can be
deduced.

However, the regions of greatest physical relevance are those at low
center of mass energy, since at high energy the rate for exclusive
production of $BB$ or $\bbpi$ is very low. Moreover, the experiments
of interest are conducted at or near threshold. Since the kinetic
energies of the $B$ and $\bbar$ are on the order of the QCD scale,
both the $B$ and $\bbar$ mesons have essentially timelike four
velocities in the center of mass frame: $v^\mu\approx
v'^\mu\approx(1,0,0,0)$.

In the second region --- beyond the resonances, but with the $B$
mesons still nonrelativistic --- the effective theory is constructed
as at high energy, but with the velocities $v$ and $v'$ equal to
$(1,0,0,0)$. As before, the virtual  photon produces a $b$ and a
$\overline{b}$ quark, each of velocity $v^\mu=(1,0,0,0)$.  In the
heavy quark theory, one can  couple a  source $S^\mu$  to the
$b$-quark current as  $\overline{b}_{v'}S^\mu\gamma_\mu {b}_v$
(i.e., with the heavy quark spins coupled to the spin of the
source). When we match onto the heavy meson theory, we can couple
the source to the mesons as $\oB(v')S^\mu\gamma_\mu\ouB(v)$. This
gives the same matrix elements as if we had coupled a source to
heavy quarks, and then evaluated the heavy quark matrix elements.

Finally, we  consider the resonance regime.  If  only hard gluons
were relevant to the binding potential,  it would be clear how to
construct such an effective theory. The matching would again proceed
in two steps.  First, one would match  onto the heavy quark theory,
and then match the $b$ quark operator onto the heavy meson effective
lagrangian. However, for a resonance which can decay into $B$
mesons, the binding is sufficiently weak that both hard and soft
gluons play a role. Hence, it might instead be appropriate to match
directly onto the low energy heavy quark chiral lagrangian.   While
we still expect that interactions with the light quarks cannot flip
a heavy quark spin, the interactions between the heavy quarks can
flip their spins.  The total spin of the heavy quarks is conserved,
however, and hence the spin of the source is transferred entirely to
the heavy  quark spins.   This can be incorporated in the heavy
meson theory by again coupling the source to the mesons as
$\oB(v')S^\mu\gamma_\mu\ouB(v)$  Fortunately, the same heavy meson
lagrangian describes the matrix elements in both scenarios, since
the diagonal subgroup of the heavy quark spin symmetry, where both
heavy quark spins are rotated simultaneously, is sufficient to
determine the form of the coupling of the heavy mesons to the
source.

The lagrangian  applicable to low-energy production of $B$ and
$\bbar$ meson is
\eqn\efflag{\eqalign{
\leff=&-i\tr\{\oB_a(v)v^\mu\partial_\mu\bB_a(v)\}
-i\tr\{\ouB_a(v)v^\mu\partial_\mu\uB_a(v)\}\cr
&+g\tr\{\oB_a(v)\bB_b(v) A^\nu_{ba}\gamma_\nu\gamma_5\}
+g\tr\{\uB_a(v)\ouB_b(v) A^\nu_{ba}\gamma_\nu\gamma_5\}\cr
&+\l_S,\cr
\l_S=&{-i\lambda\over2}S^\mu\tr\{\gamma_\mu\ouB_a(v)
\uplrarrow{D}_{ab}^\nu\gamma_\nu\oB_b(v)\}\cr
&+\lambda g' S^\mu\tr\{\gamma_\mu\ouB_a(v)
A^\nu_{ab}\gamma_\nu\gamma_5\oB_b(v)\}.}}
Here $D=\partial+V$ is the chiral covariant derivative incorporating
the pion fields. As usual, a factor of $\sqrt{M_B}$ has been
absorbed into the heavy meson fields along with the
position-dependent phase  corresponding to the momentum of the heavy
quark (so that a derivative acting on these fields only gives a
factor of the residual momentum), in order to suppress the
appearance of the heavy quark mass and emphasize the heavy quark
symmetry. Because this is the low energy theory, no large momentum
transfers are permitted. At higher energies, the appropriate
lagrangian would be the heavy meson lagrangian with velocities $v
\ne v'$. The result for two meson production matches smoothly, as
the difference in residual momenta in the amplitude gets replaced by
the difference in heavy meson velocities.

The kinetic and axial coupling terms for the $B$ mesons have been
discussed previously and result from the straightforward application
of heavy quark and chiral effective field theories \wse. ${\cal
L}_S$ is the new term and follows from the assumptions  described
above.   Note that with the trace the heavy quark spin labels are
coupled to $S^\mu\gamma_\mu$.

The coupling $\lambda$ corresponds for a crossed  process, e.g.,
$\bstr\to B\gamma$, to the Isgur--Wise form factor. Here, we treat
the ratio $\crsc{\bbpi}/\crsc{BB}$ as independent of $\lambda$,
although this is strictly true only when beyond the resonance
region. (In a full treatment of the resonance region,  $\lambda$
would have to be interpreted as a form factor with a complicated
dependence on the residual momenta, which would not drop out of this
ratio of cross sections.)
\bigbreak
\goodbreak
\xdef\secsym{5.}\global\meqno = 1
\noindent{\bf 5.\quad Calculation of
$\vect{\crsc{\three}/\crsc{\two}}{0.01}$}
\medbreak
 From the lagrangian \efflag\ we see that two types of diagrams
contribute to $\bbpi$ production. The pion can be produced
``indirectly'' by being emitted from a virtual $B$ meson through the
heavy-meson axial coupling. Or, the pion can be produced
``directly'', together with the $B$ mesons at a single vertex. This
diagram comes from  the contact term in the lagrangian in  which the
source couples directly to the $B$ meson and axial fields.  Examples
of both types of diagrams are shown in Figure~1a. We will see that
the direct contribution is much the smaller of the two, and so most
of our discussion concentrates on the indirect contribution.

In order to compare $\three$ with $\two$ as sources of neutral   $B$
mesons we normalize the $\three$ cross sections by dividing them
by the cross section for $\pair\to$ neutral $B$ mesons, $\so$,
\eqn\sigtwo{\so\equiv\crsc{B^0\bbar^0,\ B^0\bbar^{*0},
\ B^{*0}\bbar^0,\ {\rm or}\ B^{*0}\bbar^{*0}}.}
We therefore consider the ratios
\eqn\ratio{\eqalign{
\Rpp=&\sinv \sum_\pm \crsc{\three},\cr
\Rpv=&\sinv \sum_\pm
\{\crsc{BB^{*\pm}\pi^\mp}+\crsc{\bstr B^\pm\pi^\mp}\},\cr
\Rvv=&\sinv \sum_\pm \crsc{\bstr B^{*\pm}\pi^\mp},\cr}}
and also their sum,
\eqn\totalratio{R=\Rpp+\Rpv+\Rvv.}
The subscripts denote the heavy-meson content of the $\three$ final
state, with a P for each pseudoscalar and a V for each vector.   In
these ratios of cross sections, $R_\alpha$, kinematic factors
associated with the initial state and also the unknown coupling of
the source to the $B$ mesons, $\lambda$, cancel out.

The cross sections for two-body $\two$ final states are, apart from
a common factor, each given by the product of a $p$-wave
phase-space factor and a spin-counting factor \drj. The total cross
section for producing neutral $B$ mesons, $\so$, is then
proportional to a sum of such products:
\eqn\twobody{\so\propto\rpp^{3/2}+4\rpv^{3/2}+7\rvv^{3/2}\equiv
P(s).}
Here $r$ is the center-of-mass energy that remains after supplying
the rest-mass energies of the heavy-mesons. For a two-body $\two$
final state, $r^{3/2}\propto|\dkb|^3$, where $\dkb$ is the relative
three-momentum of the heavy mesons, which is the familiar $p$-wave
phase-space factor. The different values of $r$ that correspond to
the various $\two$ final states are then given by
\eqn\residual{\eqalign
{\rpp=&\rts-2M_B\cr\rpv=&\rts-M_B-M_\bstr\cr\rvv=&\rts-2M_\bstr\cr}}
For $B\bbar$ final states, the $\ra$ coincide with the $Q$ values
for the decay of the source into heavy mesons; however, we will also
use  the $\ra$, as defined above, when we consider $\three$ final
states. Note that we are including violation of the heavy-quark
symmetry as it enters through the $B$-$\bstr$ mass splitting,
$\dm=46.0\pm0.6\mev$; this splitting significantly affects the
available phase space and the off-shellness of intermediate $B$
mesons.

The density of states for a final state of $B$ mesons and a pion
simplifies for nonrelativistic $B$ mesons to
\eqn\doszero{D_0={1\over256\pi^5\epi}\,d^3\akb\,d^3\dkb
\,\delta\Bigl(r-\epi-{1\over M_B}\Bigl({1\over4}\dkb^2+\akb^2\biggr)\Bigr).}
Here $\ak^\mu=(k+k')^\mu/2$ is the average of the heavy-meson
residual momenta and $\dk^\mu=(k-k')^\mu$ is their difference.  We
have included a factor of $(\sqrt{M_B})^2$ for each of the two heavy
mesons in the final state in order to compensate for the rescaling
of the heavy meson fields in the amplitudes. If we perform the
trivial angular integrals  as well as the integral over $|\dkb|$
(using the $\delta$-function), we obtain the (integrated) density of
states,
\eqn\dosone{D={1\over16\pi^3}{M_B\over\epi}\,|\dkb|\,|\akb|^2\,d|\akb|
d(\cos\theta).}
Here $\theta$ is the angle between $\akb$ and $\dkb$. Then,
expressing $|\akb|$ and $|\dkb|$ in terms of the pion energy   and
momentum,
\eqn\subs{\eqalign{\ppib=&-2\akb
\ \Rightarrow\ |\akb|={\ppi\over2},\cr
|\dkb|=&\sqrt{4M_B\biggl(r-\epi-{\ppi^2\over 4M_B}\biggr)}
\approx2\sqrt{(r-\epi)M_B},}}
the density of states becomes
\eqn\dos{\eqalign{D=&{1\over64\pi^3}M_B|\dkb|\,|\akb|\,d\epi\,
d(\cos\theta)\cr
=&{1\over64\pi^3}M_B^{3/2}\ppi(r-\epi)^{1/2}d\epi\,d(\cos\theta).}}

As discussed in the last section, we treat the $B$ mesons as
nonrelativistic in the laboratory frame, working to lowest
nonvanishing order in the heavy-meson three-momenta. Implicit in our
approximations is the realization that the kinetic energy is fairly
evenly shared between the pion and the heavy mesons: Consider the
density of states just above (eq.\ \dos). It can be reexpressed in
terms of the total kinetic energy, $T$ ($=r-\mpi$), and the kinetic
energy of the pion, $\tpi$, as
\eqn\dosprop{\eqalign{D\propto\,&\tpi^{1/2}(T-\tpi)^{1/2}
(\tpi+\mpi)^{1/2}d\tpi\cr
=\,&T^{5/2}\,x^{1/2}(1-x)^{1/2}(x+\mpi/T)^{1/2}dx\cr}}
where $x=\tpi/T$ is the fraction of the kinetic energy given to the
pion. We find that as $T$ runs from 0 to $\gg\mpi$, the phase-space
average of $x$ runs from $1/2$ to $4/7$, i.e., on average the pion
kinetic energy about equals the sum of the $B$ meson kinetic
energies. In the differential cross section, where the density of
states is multiplied by the squared amplitude, which includes such
factors as $|\dkb|^2$ and $|\ppib|^2$, the powers of $x$ or $(1-x)$
are  increased, shifting the average value of $x$ up or down, but
the kinetic energy remains fairly evenly distributed between the
pion and the heavy mesons.

Therefore, when we work at energies where the heavy mesons are
nonrelativistic, there is a hierarchy of energy scales,
\eqn\scales{T_B,\ \epi,\ \ppi\ (\propto M_B^0)\ \ll |\k_B|
\ (\propto M_B^{1/2})\ll\ M_B,}
where $T_B$ is the kinetic energy of the $B$ mesons.  Since the
dimensionful quantities that compensate the different powers of
$M_B$ here are of order $r$, we are essentially working to lowest
order in $r/M_B$.  Taking sums and differences of the $B$ meson
momenta and using eq.\ \subs, this hierarchy can be reexpressed as
\eqn\adscales{\ak^0,\dk^0,|\akb|\ (\propto M_B^0)\ \ll|\dkb|
\ (\propto M_B^{1/2})\ll\ M_B.}
In calculating a given amplitude, we retain only the leading term
according to this hierarchy. Specifically, we drop  $\ak^0$ and
$\dk^0$ compared to $M_B$, incurring errors of order $r/M_B$.  We
also drop $|\akb|$ compared to $|\dkb|$ and $|\dkb|$ compared to
$M_B$.  This would seem to mean dropping terms of order
$\sqrt{r/M_B}$. However, once an amplitude is squared,
averaged/summed over initial/final polarizations, and integrated
over $\cos\theta$, the result depends only on $\akb^2$ and $\dkb^2$
($\akb\cdot\dkb$ vanishes in the angular integration).  Hence all
the dropped terms  are smaller by a factor of $r/M_B$.

Later, in section~8, we will consider the contribution to $\three$
production from a restricted region of phase space where the pion
energy is less than a given bound. By imposing this cutoff the
average value of the pion momentum, and therefore of $|\akb|$, will
be reduced. This only improves our approximation of neglecting
$|\akb|$ compared to $|\dkb|$.

We will illustrate these approximations in the simplest case, the
calculation of the indirect contribution to $\Rpp$.  The two graphs
that contribute are shown in Figure~1b. The graph on the left in
Figure~1b corresponds to the process $S\to B\bbst\to B(\bbar\pi)$.
Consider the propagator for the intermediate $\bbst$ state, which is
described by the velocity $v^\mu=(1,0,0,0)$ and a residual momentum
$q^\mu=k^\mu+\ppi^\mu$ and has a mass that is $\dm$ greater than
that of the $\bbar$ in the final state.  The propagator is then
\eqn\propagator{\eqalign{
{i\over2(v\cdot q+q^2/2M_B-\dm)}=&
{i\over2(v\cdot\ppi+k\cdot\ppi/M_B+\mpi^2/2M_B-\dm)}\cr\approx&
{i\over2(v\cdot\ppi-\dm)}\cr=&
{i\over2(\epi-\dm)}.}}
The first equality comes from the on-shell condition for the $\bbar$
meson, $(v+k/M_B)^2=1$, which is just the nonrelativistic formula
$T_B\approx\k^2/2M_B$. The next line results from  neglecting terms
of order $k\cdot\ppi/M_B$ and $\mpi^2/M_B$ compared to $\epi$, in
accordance with the hierarchy in eq.\ \adscales. The propagator we
have obtained, which is inversely proportional to the difference in
energy between the intermediate and final states, is simply that
prescribed by nonrelativistic, time-ordered perturbation theory.

Now consider the part of the amplitude coming from the $S\to B\bbst$
vertex. From our effective lagrangian \efflag\ we see that it is
proportional to $\q-\k'=\dkb+\ppib$. Then according to the hierarchy
\adscales, we can drop $\ppib=-2\akb$. The amplitude for $S\to
B\bbst\to B(\bbar\pi)$ is then  given by
\eqn\amp{\a_1(\e;v,k,k')\approx-{ig\lambda\sqrt{2}\over2(\epi-\dm)\fpi}
\eabcd v^\alpha \dk^\beta \ppi^\gamma \e^\delta,}
where $\e$ is the (purely spatial) polarization of the source. The
second graph, shown on the right in Figure~1b, is related to the
first by charge conjugation and isospin, so that their sum is given
by
\eqn\sumamp{\eqalign{\a&=\a_1(\e;v,k,k')-\a_1(-\e;v,k',k)
\approx2\a_1(\e;v,k,k')\cr
&\approx-{ig\lambda\sqrt{2}\over(\epi-\dm)\fpi}
(\dkb\times\ppib)\cdot\eb.}}
Squaring and summing over the averaging over initial polarizations
we obtain
\eqn\sqamp{\langle|\a|^2\rangle_\e\approx{2\over3}
\biggl({g\lambda\over(\epi-\dm)\fpi}\biggr)^2\dkb^2\ppib^2(1-\cos^2\theta).}
If we substitute for $|\dkb|$, integrate over the density of states,
and divide by $\so$ (not including the common kinematical factors
corresponding to the initial state) we obtain $\Rpp$.

The cross-section ratios for the various $\three$ final states are
given by
\eqn\ratioresults{
\Rai(s)={2g^2\over 3\pi^2\fpi^2}{1\over P(s)}
\int_{\mpi}^{\ra} d\epi \ppi^3 (\ra-\epi)^{3/2}
\times\cases{{1\over(\epi-\dm)^2},&$\alpha$=PP;\cr
\noalign{\vskip4pt}
{7/4\over(\epi-\dm)^2}+{1/2\over\epi^2-(\dm)^2}\cr
\quad\quad\,\,+{1\over\epi^2}+{3/4\over(\epi+\dm)^2},&$\alpha$=PV;\cr
\noalign{\vskip4pt}
{5\over\epi^2}+{2\over(\epi+\dm)^2},&$\alpha$=VV.\cr} }
Here the $i$ signifies that these are the indirect contributions to
$\Ra$. The upper integration limit is the value of $r$ appropriate
to the heavy-meson content of the $\three$ final state, as given by
eq.\ \residual.

At a fixed center of mass energy, the pion attains its maximum
energy when the heavy meson pair is produced with zero relative
momentum ($\dk=0$), back to back with the pion.  From eq.\ \subs\ we
see that, to the accuracy we are are working to, this occurs when
$\epi=\ra$. Hence the upper limit of the phase-space integral over
pion energy is $\ra$.

As already discussed, there is  also a direct vertex that describes
the production of $B$ mesons and a pion at a single vertex.  For
$\three$ final states that include at least one heavy vector, it is
possible to produce the heavy mesons in an $s$ wave. Then, the
amplitudes are proportional to $\epi/(M_B\fpi)$ while the amplitudes
for the indirect graphs considered above are proportional to
$|\dkb|\ppi/(M_B\fpi\epi)$.  At large $r$, where the cross section
is dominated by contributions with $\epi\gg\mpi$, the ratio of
direct and indirect amplitudes is proportional to $\epi/|\k_B|$
which is of order $\sqrt{r/M_B}$.  Therefore, the $s$-wave
contributions of the direct graphs are suppressed by a factor of
$r/M_B$, which is small for $r$ inside the region of validity of the
chiral expansion. The $s$-wave direct graphs, which cannot interfere
with the $p$-wave indirect graphs, contribute to the $R_\alpha$ as
\eqn\directratioresults{
R^d_\alpha(s)={3g'^2\over 2\pi^2\fpi^2}{1\over P(s)}
\int_{\mpi}^{\ra} d\epi \epi^2 \ppi (\ra-\epi)^{1/2}
\times\cases{
0,&$\alpha$=PP;\cr
\noalign{\vskip4pt}
1/M_B,&$\alpha$=PV;\cr
\noalign{\vskip4pt}
1/M_B,&$\alpha$=VV.\cr} }
The $d$ stands for direct. Note that the source cannot decay into
two pseudoscalars in an $s$-wave and a pion without violating either
angular momentum or parity conservation.  Hence $\Rpp^d$ is zero to
this order in $1/M_B$.
%
\def\sp{$\ $}
\def\row#1#2#3#4#5#6#7#8#9
      {&$#1$&&$#2$&&$#3$&&$#4$&&$#5$&&$#6$&&$#7$&&$#8$&&$#9$&\cr}
\def\rowomit{height2pt
      &\omit&&\omit&&\omit&&\omit&&\omit&&\omit&&\omit&&\omit&&\omit&\cr}
\def\horizrule{\rowomit\noalign{\hrule}\rowomit}
\topinsert
\centerline{Table 2. The ratios
$\crsc{\three}/\crsc{\two}$ expressed as percentages.}
$$\vbox{\offinterlineskip
\hrule\halign{
\vrule#&\sp\hfil#\hfil\sp& \vrule#&\sp\hfil#\hfil\sp&
\vrule#&\sp\hfil#\hfil\sp& \vrule#&\sp\hfil#\hfil\sp&
\vrule#&\sp\hfil#\hfil\sp& \vrule#&\sp\hfil#\hfil\sp&
\vrule#&\sp\hfil#\hfil\sp& \vrule#&\sp\hfil#\hfil\sp&
\vrule#&\sp\strut\hfil#\hfil\sp&\vrule#\cr
\rowomit
\row{\sqrt{s}({\rm MeV})}{\Rpp^i}{\Rpv^i}{\Rvv^i}{R^i}{\Rpv^d}{\Rvv^d
}{R^d}{R}
\horizrule
 \row{10865}{0.12}{0.12}{0.02}{0.26}{0.03}{0.01}{0.04}{0.30}
 \row{11020}{0.53}{0.99}{0.70}{2.2 }{0.19}{0.10}{0.29}{2.5 }
 \row{11200}{1.3 }{3.1 }{3.1 }{7.6 }{0.59}{0.40}{0.99}{8.6 }
 \row{11500}{3.3 }{9.2 }{11. }{24. }{2.0 }{1.6 }{3.7 }{27. }
 \row{12000}{8.2 }{26. }{36. }{70. }{7.8 }{6.7 }{14. }{85. }
\rowomit
\noalign{\hrule}
}}$$
\endinsert

The axial coupling of the heavy mesons,  $g$, has been bounded above
by $g^2\le0.5$ \ref\amd\amdref{J.~Amundson, C.~G.~Boyd, E.~Jenkins,
M.~Luke, A.~Manohar, J.~Rosner, M.~Savage, M.~Wise,  UCSD/PTH
92--31, hep--ph/9209241.} using the experimental upper limit for the
$D^{*+}$ width \ref\acc\accref{ACCMOR Collaboration. S.~Barlag, et.\
al.\ Phys.\ Lett.\ B278 (1992) 480.}  which is dominated by
$D^{*+}\to D^0\pi^+$ and $D^{*+}\to D^+\pi^0$. Using the maximum
allowed value of $g^2$ and taking $g'^2=1$,   the numerical values
of the ratios $\Ra$ (expressed as percentages) are given in Table~2
for various values of the center-of-mass energy, $\sqrt{s}$, and are
plotted in Figure~2. The sum of the direct contributions to $R$ is
also displayed in Figure~2. Because they are so small, we neglect
them in the rest of our discussion. The overall sum is uncertain due
to the uncertainties in both the direct and indirect contributions.

At the $\usx$ resonance, the $B$ mesons are produced with a charged
pion only a few percent as often as they are produced alone.   Below
this resonance, for example at the $\ufv$, this ratio is negligible,
but above it grows with increasing energy as the limited three-body
phase space for pion production is overcome. For center of mass
energy of $12\gev$, the rate is almost comparable to the rate
without a pion.

Returning to the results for the $\Rai$ given in eq. \ratioresults,
we see that in the limit of exact heavy quark symmetry, $\dm=0$, the
ratios $R_\alpha$ are proportional to one another as
\eqn\proportionality{\Rpp^i:\Rpv^i:\Rvv^i::1:4:7\qquad(\dm=0).}
This limit is approximately realized for $\dm\ll r\ll M_B$, where
the effects of $\dm\neq0$ are minimized and the nonrelativistic
treatment still applies.     For such large values of $r$, it is
also a good approximation to take $\mpi=0$. Then we find
$\Rai\propto r^{7/2}$.  The two extra powers of $r$ relative to the
$r^{3/2}$ scaling of the $\two$ cross sections reflect the
derivative coupling and additional phase space factor for the pion.

The proportions $1:4:7$ are the same as was found for the production
rates of heavy mesons without an accompanying pion in the various
spin states  (PP, PV and VV) using simple spin counting \drj\ (see
eq. \twobody). We can use the same method to understand the
persistence of the proportions $1:4:7$ when a pion is emitted from
one of the heavy mesons.

In $S\to B\bbar$ (pseudoscalars and vectors),  the heavy quark spins
are fixed so as to carry the spin of the source.  Charge-conjugation
invariance then determines the spin state of the light quarks:
Consider the $B\bbar$ final state from the point of view of heavy
quark symmetry,  with each heavy meson described by a bispinor.
Under charge conjugation, the positions and the heavy- and
light-quark spins labels of the $B$ and $\bbar$ mesons  are
interchanged.  Since the heavy mesons are in a $p$ wave and the
heavy quark spin state is symmetric (spin one), the light quark spin
state must also be symmetric if the final state is to be
charge-conjugation odd like the source.  Hence the total spin of the
light quarks must be one.   Because the spin of the source is
carried by the heavy quarks, after the angular positions of the
heavy mesons have been integrated over (but with the source spin
fixed), the total light quark spin points with equal probability in
all directions.  Adding the light and heavy quark spins in each
heavy meson to find the meson spins, one finds the ratios $1:4:7$.

Now consider the case where a pion is emitted from a heavy meson.
Because of isospin invariance, we need only consider the case where
the pion is neutral and hence self-conjugate. Since the $\pi^0$ is
charge-conjugation even, the total spin of the light quarks in the
heavy mesons must still be one. After integrating over the angular
positions of all the particles,  the light quark spin distribution
will again be isotropic.  Hence in the heavy quark limit, the heavy
meson spin proportionalities do not change when a pion is emitted
from one of the heavy mesons.
\bigbreak
\goodbreak
\xdef\secsym{6.}\global\meqno = 1
\noindent{\bf 6. \quad Implications for CP Violation Studies}
\medbreak
We have calculated  the $\bbpi$ branching ratio primarily in order
to assess the prospects for studying CP violation in the
$B^0$-$\bbar^0$ system using Yamamoto's pion-tagging method \yam.
Let us now see if the $\three$ branching ratios obtained in the
previous section are large enough to produce the statistics required
to resolve the small, CP-violating asymmetries in the decays of
neutral $B$ mesons. The very small values found for $R$ at the
$\ufv$ and $\usx$ resonances indicate that the pion-tagging method
will not be useful in the resonance regime, but the  rapid increase
in the rate with center of mass energy suggests that the method
could prove useful at higher energies, above the resonances.

The analysis of this section is due to Yamamoto \yam, but uses our
results for the $\three$ branching ratios.  We follow him in
introducing a figure of merit which measures the statistical power
of a given method of measuring CP violation,
\eqn\fom{f=\sigma\et d^2.}
It is proportional to  the cross section, $\sigma$, for the process
in which the $B$ mesons are produced. There is a factor of the
tagging efficiency, which is taken to include  the number of neutral
$B$ mesons produced in the process. Finally, it includes the square
of the ``dilution factor'', $d$, which is given by
\eqn\acpdef{\acp=d\sin2\beta,}
where $\beta$ is a CP-violating angle that appears in the
CP-violating asymmetry, $\acp$, measured in  neutral $B$ decay.  For
example, using the pion-tagging method one measures the asymmetry
\eqn\acpex{\acp_{\three}
={N(B\to\pk)-N(\bbar\to\pk)\over N(B\to\pk)+N(\bbar\to\pk)}
={x\over1+x^2}\sin 2\beta,}
where $N$ is the number of events from a given decay process,
$x=\delta M/\Gamma\approx0.73$ ($\delta M$ is the mass difference
between the CP-even and -odd linear combinations of $B^0$ and
$\bbar^0$ and $\Gamma$ is their common lifetime),  and $\pk$ is a
representative CP eigenstate. The utility of the figure of merit is
that it is inversely proportional to the integrated luminosity
required in order to measure a CP-violating angle; the larger the
figure of merit, the easier it is to measure the angle to a given
precision.

In Table~3 we compare the figure of merit for the pion-tagging
method, at the $\usx$ and at $12\gev$, with the familiar methods of
generating neutral $B$ mesons for CP-violation studies: $\ufr\to
B^0\bbar^0$ at an asymmetric collider and $\ufv\to B^0\bbar^0\gamma$
at a symmetric collider.  Although the asymmetric collider provides
the best figure of merit, it is worthwhile to determine alternative
methods at a symmetric collider in case either such a machine is not
constructed or the required luminosity is not obtained. The relevant
benchmark for us is therefore comparison with the figure of merit
for studies using $\u\to B\bbar\gamma$, which are feasible at a
symmetric collider.

The $BB(X)$ cross section is approximately $0.3\nb$ at the $\ufv$
and $\usx$ resonances. Using the value of $R=2\%$ at the $\usx$
found above, not surprisingly we find that the pion-tagging method
is not competitive at the $\usx$.

To estimate the $\three$ cross section at $12\gev$, beyond the
resonances, we take the $(BB)X$ cross section as  $1/2$ of its
on-resonance value at the $\usx$.   We further assume that $(BB)X$
is dominated by  pair production of strange and nonstrange $B$
mesons and $B$ mesons accompanied by a pion (charged or neutral).
We expect that because of the sharing of kinetic energy discussed
earlier, multipion production will be much smaller than single pion
production at this energy. To estimate the contribution of $BBK$ and
$BB\eta$ production to $(BB)X$, we repeated our calculation of
$\three$ production above but with $K$ and $\eta$ mesons radiated
from the heavy mesons, and including SU(3) breaking via the $K$ and
$\eta$ masses and decay constants and the $B_s$-$B$ mass splitting.
We found that $BBK$ and $BB\eta$ together are about 1/4 $\bbpi$, so
we neglect their contribution to (BB)X.  Estimating $B_s$ meson
production, as discussed in the following section, by using eq.\
\twobody\ but with the values of $\ra$ determined by the $B_s$ meson
masses, we find that $B_s$ mesons are produced about 80\% as often
as $B^0\bbar^0$. We therefore estimate that the $\three$ cross
section is $\approx0.15\nb\times R/(2.8+3R/2)\approx0.027\nb$.  We
find a figure of merit of $0.005$.

If we extrapolate our results for $\three$ production to even higher
energies, we find that at about $12.5\gev$, $R$ is about twice its
value at $12\gev$, so that the figure of merit is also roughly
doubled.  We conclude that
the figure of merit can be comparable to the figure of merit
for $\ufv\to B^0\bbar^0\gamma$ at a symmetric collider.
Because our calculation is not exact, it is conceivable
that the figure of merit is even higher.   This can only be
determined experimentally.

We emphasize that it is unlikely that we have vastly overestimated
the rate, since it is clear that as one
approaches higher center of mass energy that pion production will be
less suppressed. So long as the center of mass energy is not too
high, single pion production will dominate. Moreover, we
show in section 8 that even if we restrict our integral over the phase space to
pion
energies for which the chiral results are certainly reliable, we
still get a substantial fraction  the total cross section determined
from the full phase space integration.

We conclude that Yamamoto's pion-tagging method
should be competitive.  An advantage of running at a higher center of mass
energy would be that it could be possible to use both $B B^*$
production and $\three$ production simultaneously. This might
augment statistics, and provide a useful check on both methods.
%
\def\row#1#2#3#4#5
      {&#1&&$#2$&&$#3$&&$#4$&&$#5$&\cr}
\def\rowomit{height2pt
      &\omit&&\omit&&\omit&&\omit&&\omit&\cr}
\def\horizrule{\rowomit\noalign{\hrule}\rowomit}
\topinsert
\centerline{Table 3. Comparison of the pion-tagging method with standard}
\centerline{methods of studying CP violation in the $B$ system.}
$$\vbox{\offinterlineskip
\hrule\halign{
\vrule#&\quad#\hfil\quad&
\vrule#&\quad#\hfil\quad&
\vrule#&\quad\hfil#\quad&
\vrule#&\quad\hfil#\hfil\quad&
\vrule#&\quad\strut#\hfil\quad&\vrule#\cr
\rowomit
\row{\hfil Mode}{\hfil\sigma}{\et\hfil}{d}{\hfil\sigma\et d^2}
\horizrule
\rowomit
\row{$B^0\bbar^0$ at $\ufr$}{0.5\nb}{2\times0.4}{x/(1+x^2)}{0.092}
\row{$B^0\bbar^{0}\gamma$ at $\ufv$}{0.05\nb}{2\times0.4}{2x/(1+x^2)^2}{0.015}
\row{$\three$     at $\usx$}{0.004\nb}{0.8}{x/(1+x^2)}{0.0006}
\row{$\three$     at $12\gev$}{0.0027\nb}{0.8}{x/(1+x^2)}{0.005}
\rowomit
\noalign{\hrule}
}}$$
\endinsert
\bigbreak
\goodbreak
\xdef\secsym{7.}\global\meqno = 1
\noindent{\bf 7. \quad Implications for $\vect{B_s}{0.01}$ Identification}
\medbreak
Two methods have been used for identifying $B_s$ mesons at a
resonance:  One is to scan about the resonance and look for an
increased production of strange mesons.  Another method is to
identify a $B_s$ meson by the lower endpoint of the spectrum of a
lepton produced in its decay; $B_s$ mesons are heavier than
nonstrange $B$ mesons and are therefore produced with smaller
momenta; consequently, a lepton from $B_s$ decay will have smaller
momentum than a lepton from $B$ decay. However, nonstrange $B$
mesons produced with a pion also have less momentum than those
produced alone, and therefore $BB\pi$ is a potential source of
background for $B_s$ identification.

We normalize $B_s$ production to nonstrange $B$ meson production,
\eqn\rs{R_s\equiv\crsc{B_s\bbar_s,\ B_s\bbar^*_s,
\ B^*_s\bbar_s,\ {\rm or}\ B^*_s\bbar^*_s}/\so,}
where $\so$ is the cross section for production of neutral,
nonstrange $B$  mesons of spin 0 and 1 defined above in eq. \sigtwo.
If we assume that the coupling of the source to the heavy mesons is
SU(3) symmetric, then up to a common factor both the strange and
nonstrange cross sections are given by eq. \twobody\ but with
different values of $\ra$ because of the $B_s$-$B$ mass difference.
(The hyperfine splittings of the strange and nonstrange $B$ mesons
are essentially equal, however:  $M_{B^*_s}-M_{B_s}\approx
M_{B^*}-M_B$.)  To determine the importance of the $\bbpi$
background we compare $R_s$ with $3R/2$; $R$ must be multiplied by
3/2 in order to include $\bbpi^0$ production.

Because of the large uncertainty in the experimental value for the
strange-nonstrange mass splitting, between 80 and $130\mev$
\ref\lee\leeref{CUSB Collaboration. J.~Lee-Franzini et.~al., Phys.\
Rev.\ Lett.\ 65, 2947 (1990).},  we calculate $R_s$ for three
different values of the splitting that span the allowed range: 80,
105, and 130$\mev$.  The results, expressed as percentages,  are
shown for center of mass energies corresponding to the $\ufv$ and
$\usx$ resonances in Table~4.
%
\def\row#1#2#3#4#5
      {&$#1$&&$#2$&&$#3$&&$#4$&&&$#5$&\cr}
\def\rowomit{height2pt
      &\omit&&\omit&&\omit&&\omit&&&\omit&\cr}
\def\horizrule{\rowomit\noalign{\hrule}\rowomit}
\topinsert
\centerline{Table 4. Comparison of $R_s$ (in \%), for various values of
$M_{B_s}-M_B$, with ${3\over2}R$}
$$\vbox{\offinterlineskip
\hrule\halign{
\vrule#&\quad\hfil#\hfil\quad&
\vrule#&\quad\hfil#\quad&
\vrule#&\quad\hfil#\quad&
\vrule#&\quad\hfil#\quad&
\vrule#&$\,$
\vrule#&\quad\strut\hfil#\quad&\vrule#\cr
\rowomit
\row{\rts}{80\mev}{105\mev}{130\mev}{{3\over2}R\,(\%)}
\horizrule
\rowomit
 \row{\ufv        }{20.}{5.7}{.76}{.39}
 \row{\u(6{\rm S})}{46.}{32.}{20.}{3.3}
 \rowomit
\noalign{\hrule}
}}$$
\endinsert
At the $\ufv$ resonance, just above $B_s\bbar_s$ threshold,  we find
that the $B_s\bbar_s$ cross section depends strongly on the
$B_s$-$B$ mass splitting which determines the size of the $p$-wave,
$B_s$-$\bbar_s$ phase space. As a result, the relative size of the
$\bbpi$ background also depends strongly on this splitting.  For the
largest allowed value of the $B_s$-$B$ mass splitting, the $\bbpi$
cross section is about 1/2 of the $\twos$ cross section while at low
end it is only a few percent.  From
the application of heavy-quark symmetry at leading order, we  expect
the $B_s$-$B$ mass difference to be close to the $D_s$-$D$ mass
difference.  This  is measured as $99.5\pm0.6\mev$  \pdg, not far
from the center of the allowed range of the $B_s$-$B$ splitting. At
this point $BB\pi$ is about $10\%$ of $B_s\bbar_s$.  At the $\usx$,
the  $B_s$-$B$ mass splitting is less important, and the $\bbpi$
background is of order $10\%$ of the $B_s\bbar_s$ cross section. We
conclude that the $BB\pi$
background is probably not a  problem for $B_s$ identification at either
of the $\ufv$ and $\usx$ resonances.

Of course, it should be borne in mind that this calculation assumed
SU(3) symmetry and was based on heavy meson effective chiral theory
applied in the resonance regime.  However, this naive calculation
indicates that $\bbpi$ background should not be a problem.
\bigbreak
\goodbreak
\xdef\secsym{8.}\global\meqno = 1
\noindent{\bf 8.\quad Measurement of $\vect{g}{0.01}$}
\medbreak
So far, we have integrated over the entire phase space because both
CP violation studies and the identification of $B_s$ mesons rely on
the total rate. In this section, we instead focus on the region of
phase space where we expect the calculation to be reliable.  This
serves two purposes. First, it allows for the possible extraction of
the axial coupling constant of the $B$ mesons, $g$.  It would be
useful to directly extract $g$ in this way and compare to the bounds
on $g$ in the $D$ system. This method is probably not useful for
extracting $g$ in the $D$ system however as we show below. Second,
we can establish a reliable lower limit on the branching fraction to
$BB\pi$ for center of mass energies beyond the resonance regime.

Because we want a reliable prediction, we focus on energies beyond
the resonance region. We also need to determine when the derivative
expansion of the heavy quark chiral lagrangian is sufficiently
reliable that we can trust the leading order (in derivatives)
result. As stated in ref.\ \wse, we need both $v\cdot\ppi$ and
$v'\cdot\ppi$ to be small. How small depends on the cutoff for the
theory; this may be the chiral symmetry breaking scale, $\csb$ or it
may be smaller (see ref.\ \ref\qcd\qcdref{ Lisa Randall and Eric
Sather, MIT-CTP-2167, Nov.~1992.  $\langle$Bulletin Board:
hep-ph@xxx.lanl.gov - 9211268$\rangle$.} for a discussion). We call
the cutoff $\Lambda$.

Notice that in order for both $v\cdot\ppi$ and  $v'\cdot\ppi$ to be
small, $v\cdot v'$ and $v^0$ are restricted. We see this by adding
both constraints together, which yields $\av^0\epi< \Lambda$. The
lowest possible $\epi$ is $\mpi$, implying $\av^0<\Lambda/\mpi$
which in turn implies $v \cdot v' <2(\Lambda/\mpi)^2-1$.

This means that even for the pion emitted at threshold, the momentum
of the heavy mesons is constrained if the chiral expansion is to be
valid, and furthermore, that the best region to apply the heavy
meson lagrangian will be in the region where $\av^0 \approx 1$ but
above the resonance region, where the calculation will be valid over
the maximum possible range of pion energies.  Therefore, we will
concentrate on extracting $g$ in the regime where the $B$ mesons are
nonrelativistic. This is clearly the best place from an experimental
vantage point, and as we have argued, is probably also the region
where the branching fraction to $B$ mesons and a single pion is
maximal. One might also hope to extract $g$ for $D$ mesons. However,
at CLEO, where $D$ mesons are copiously produced, $v^0_D \approx 3$.
Therefore, even with a high cutoff for $\Lambda$ of order $1\gev$,
one could only integrate to pion energies less than $2\mpi$. Even
with this limited region of phase space, $v\cdot\ppi$ and $v'\cdot
\ppi$ are very close to the cutoff so that the extraction of $g$ for
$D$ mesons is probably not reliable. So we focus on the extraction
of $g$  in the $B$ system.

Recall that our expression for $R$ --- the ratio of the cross
section for  production of the various $\three$ final states divided
by the cross section for $B\bbar$ production --- was expressed as an
integral over pion energies (see eq.\ \ratioresults).  We now
integrate the differential cross section over $\epi$ only up to some
maximum pion energy, $\emax$.  We will refer to the result as
$\Rmax$.  Ultimately one wants to choose $\emax$ as the maximum pion
energy for which we expect the cross section to be reliable. The
allowed range of $\emax$ is between $\mpi$ and $\rpp$.

In Figure~3, we plot $\Rmax$ as a function of $\emax$ for  $\rts$ of
$11.5\gev$, $11.75\gev$ and $12\gev$.  Note that for
$\emax\lapprox500\mev$, $\Rmax$ is roughly independent of $\rts$ in
this range. This is because at these energies,  the $p$-wave
phase-space  factor for the heavy mesons is not much different for
heavy mesons produced with a soft pion than for heavy mesons
produced alone, and approximately cancels out in the ratio $\Rmax$.
Accordingly, for $\emax=400\mev$ we find that $\Rmax$ is roughly
10\% for any value of $\rts\gapprox11.5\gev$; for $\emax=500\mev$ we
find  $\Rmax$ slowly varies from about 15\% to 20\% as $\rts$ runs
from $11.5$ to $12\gev$.

For larger values of $\emax$, the dependence of $\Rmax$ cannot be
ignored. If we allow a large value of the cutoff, $\emax=800\gev$,
we find that we gain rather little at $\rts=11.5\gev$, moving up to
a value of $\Rmax$ just above 20\%, whereas at $11.75\gev$ we have
increased to over 35\%, and at $12\gev$ to over 45\%.

This analysis also allows us to conclude that we can reliably
predict a large value of $R_{\alpha}$, since within the regime of
reliability of our calculation, the ratio $R_{\alpha}$ is very
large, greater than 45\% at $\rts=12\gev$  if we allow a cutoff of
$800\mev$. It is encouraging that this rate is so large.

It is important to recognize that one can first test that the heavy
quark symmetry relations apply, by studying the relations among
pseudoscalar and vector $B$ meson production without a pion. If
these prove valid, one can then proceed to measure the cross section
with a sufficiently small cut on the pion energy so that the chiral
theory is applicable. By varying the cutoff, one could  test where
the results disagree with our prediction, indicating that the
derivative expansion of the heavy meson effective theory has broken
down. This should permit a reliable extraction of $g$.
\bigbreak
\goodbreak
\xdef\secsym{9.}\global\meqno = 1
\noindent{\bf 9.\quad Conclusion}
\medbreak
It is clear that the proposal of Yamamoto is quite interesting.  It
appears that there might be a sufficiently large rate for
self--tagging $B$ meson events  for this to be a viable method of
studying CP violation for neutral mesons at a symmetric collider.
Despite the  limitations of our calculation, we can nevertheless
establish several interesting results.  Within the resonance regime,
the process we consider will probably not occur at a sufficiently
large rate to compete with the more conventional proposal for the
study of CP violation at a symmetric collider, namely $BB^*$
production followed by $B^*\to B\gamma$.   At center of mass energy
of about $12000\mev$, pion emission from a $B$ meson pair could
occur as often as not. Although the calculation here is reliable
only  over about half the range of pion energies, it is clear that
the rate for a single pion accompanying the $B$ mesons is large,
even from this restricted phase space.

It could therefore be useful to run at center of mass energy above
the resonance region. In this region, one has the advantage that  $B
B^*$ and $\three$ production should both be large. With two
different methods of looking for CP violation, more reliable results
might be obtained. And of course, the $\three$ process has a much
cleaner tag and employs known technology.

Furthermore, it is likely that we have been conservative in our
evaluation of the utility of Yamamoto's method. We assumed a drop in
cross section by a factor of 2 at center of mass energy $12\gev$. In
addition, we have neglected production of resonant $B$ meson states
which would decay primarily to $B\pi$.  Because the production of
such  a resonance with a low-lying $B$ meson state is not $p$-wave
suppressed, there could be a  larger rate for $\three$ than we
computed.

 From our calculation, we have also established that at the  $\ufv$,
the production of $\bbpi$ is probably not a large background.

Finally, we have shown that the rate for $\three$ outside the
resonance regime should allow for a reliable extraction of the
axial pion heavy meson coupling constant, $g$.  In general, running
above resonance but in the regime where $B$ mesons are
nonrelativistic could be the best place from the point of view of
testing heavy quark relations at energies sufficiently high that
they should be reliable, but not so large that exclusive modes are
suppressed. It is encouraging that the rate is largest in the range
of energies where the calculation should be most reliable.
\bigbreak
\bigbreak
\centerline{\bf Acknowledgements}
\medskip
\nobreak
{\narrower We thank Hitoshi Yamamoto and Howard Georgi  for
motivating this work and for  valuable discussions. We are grateful
to Hitoshi  for answering our many experimental questions. We also
thank Mark Wise for suggesting the extraction of $g$. \smallskip}
\vfill
\eject
\baselineskip 12pt plus 1pt minus 1pt
\centerline{\bf References}
\bigskip
\item{\yam}\yamref
\medskip
\item{\geo}\georef
\medskip
\item{\gri}\griref
\medskip
\item{\cwz}\cwzref
\medskip
\item{\wse}\wseref
\medskip
\item{\pdg}\pdgref
\medskip
\item{\jaf}\jafref
\medskip
\item{\dug}\dugref
\medskip
\item{\rep}\repref
\medskip
\item{\man}\manref
\medskip
\item{\flk}\flkref
\medskip
\item{\drj}\drjref
\medskip
\item{\dex}\dexref
\medskip
\item{\dgg}\dggref
\medskip
\item{\ley}\leyref
\medskip
\item{\amd}\amdref
\medskip
\item{\acc}\accref
\medskip
\item{\lee}\leeref
\medskip
\item{\qcd}\qcdref
\vfill
\eject
\centerline{\bf Figure Captions}
\bigskip
\item{Figure~1.}Diagrams that contribute to $\bbpi$ production. The
blob represents the source $S^\mu$. (a) Examples of the two kinds of
graph: indirect and direct. (b) The two indirect graphs that
contribute to the production of two pseudoscalar $B$ mesons and a
pion.
\medskip
\item{Figure~2.}The ratio, $R$, of the $\three$ and $B^0\bbar^0$
cross sections as a function of the center of mass energy.  Both
indirect and direct contributions are shown. Also shown are the
individual indirect contributions, $\Rai$, coming from specific
$\three$ final states.  The couplings were taken as $g^2=0.5$ and
$g'^2=1$.
\medskip
\item{Figure~3.}The (indirect) contribution to $R$ for pion energies
less than a cutoff $\emax$ for center of mass energies of $11.5$,
$11.75$ and $12\gev$.
\vfill
\eject
\end